# Independently switchable atomic quantum transistors by reversible contact reconstruction

F.-Q. Xie[1,4*], R. Maul[2*], A. Augenstein[1], Ch. Obermair[1,4], E.B. Starikov[2],

G. Schön[2,3,4], Th. Schimmel[1,2,4], W. Wenzel[2,4],

[1] Institut für Angewandte Physik, Universität Karlsruhe, 76128 Karlsruhe, Germany

[2] Forschungszentrum Karlsruhe, Institut für Nanotechnologie, 76021 Karlsruhe, Germany

[3] Institut für Theoretische Festkörperphysik, Universität Karlsruhe, 76128 Karlsruhe, Germany

[4] DFG-Center for Functional Nanostructures (CFN), Universität Karlsruhe, 76128 Karlsruhe, Germany

[*] authors contributed equally

**The controlled fabrication of actively switchable atomic-scale devices, in particular transistors, has remained elusive to date. Here we explain operation of an atomic-scale three-terminal device by a novel switching mechanism of bistable, self-stabilizing reconstruction of the electrode contacts at the atomic level: While the device is manufactured by electrochemical deposition, it operates entirely on the basis of mechanical effects of the solid-liquid interface. We analyze mechanically and thermally stable metallic junctions with a predefined quantized conductance of 1-5 $G_0$ in experiment and atomistic simulation. Atomistic modelling of structural and conductance properties elucidates bistable electrode reconstruction as the underlying mechanism of the device. Independent room-temperature operation of two transistors at low voltage demonstrates intriguing perspectives for quantum electronics and logics on the atomic scale.**

Controlling the electronic conductivity on the quantum level will impact the development of future nanoscale electronic circuits with ultra-low power consumption. Fascinating physical properties and technological perspectives have motivated intense investigation of atomic-scale metallic point contacts in recent years[1-10]. The quantum nature of the electron is directly observable, because the width of the contacts is comparable to the electron wavelength and conductance is quantized in multiples of $2e^2/h$ in ideal junctions. In real metallic point contacts, which have been fabricated by mechanically controlled deformation of thin metallic junctions[1, 11-13] and electrochemical fabrication techniques[9], the conductance depends on the chemical valence[3]. Two-terminal conductance-switching devices based on quantum point contacts were developed with an STM-like setup[4] and electrochemical methods[5]. Recently quantized magnetoresistance in atomic-size contacts was switched between two quantized conductance levels by rotating the point contacts in a magnetic field[6].

We have developed a three-terminal gate-controlled atomic quantum switch with a silver quantum point contact in an electrochemical cell, working as an atomic-scale relay[7]. We control individual atoms in a quantum point contact by an independent gate electrode, which allows a reproducible switching of the contact between a quantized conducting "on" state and an insulating "off" state(Fig 1a). Fig. 1b demonstrates long sequences of electrochemically controlled switching between the non-conducting "off-state" and the quantized conducting "on-state", where the quantum conductance of the switch follows the gate potential, as commonly observed in transistors.

When we set the gate potential to an intermediate "hold" level between the "on" and "off" potentials, the currently existing state of the atomic switch remains stable and no further switching takes place. This is demonstrated in Fig. 1c both for the "on-state" of the switch (left arrow) and for the "off-state" of the switch (right arrow). Thus the switch can be reproducibly operated by the use of three values of the gate potential for "switching on", "switching off" and "hold". This provides the basis for atomic-scale logical gates and atomic-scale digital electronics.

The smallest observed conductance value was 1 $G_0$, but we have significantly optimized our protocol to configure quantum conductance switches with any integer multiple $1 \leq n \leq 5$ of $G_0$ (higher values have also been observed) at will (see Figure 2a). While we can understand the conductance properties of such junctions on the basis of atomistic conductance calculations[8] the physical process underlying the switching mechanism remained unclear. Reproducible switching between quantum-conductance levels over many cycles cannot be explained by conventional atom-by-atom deposition, but requires a collective switching mechanism. Our previous calculations have shown that only well-ordered junction geometries result in integer multiples of the conductance quantum. Neither partial dissolution of the junction nor its controlled rupture yields the necessary atomic-scale memory effect. A more detailed model of the structural[11, 12, 14, 15] and conductance [3],[16] properties of such junctions is therefore required.

We begin with the unbiased deposition of silver ions (Fig 3(a)), which evolve under the influence of the electrostatic potential of the electrodes[17]. Starting from two distant parallel Ag(111) layers we evolve each ion in the long-range electrostatic potential generated from the present electrode conformation and a short-range Gupta potential for silver[18] (see suppl. material). In the simulations we deposit one atom at a time using a kinetic Monte-Carlo Method[17]: starting at a random position inside the junction. We deposit up to 800 atoms in the junction until a predefined number of non-overlapping pathways connect the left and right electrode. As non-overlapping pathway we define a unique set of touching atoms that extend from one electrode to the other, which permits to identify the minimal cross-section of the junction.

Next, we simulate the switching process: The change in the electrochemical potential induces a change in interface tension of the liquid-metal interface, making possible a deformation of the junction geometry parallel the junction axis. It is well known that changes in the electrochemical potential modulate the interfacial tension of the whole electrode[19-21], which results in a mechanical strain on the junction. We simulate the opening/closing cycle of a junction by evolving the atoms of a "central" cluster under the influence of the electrochemical pressure. We assume that only the atoms in this region move in the switching process, while most of the bulk material remains unchanged. The central cluster comprises the atoms of the minimal cross-section connecting the two electrodes and all atoms within a radius of 9.0 Å around this central bottleneck. While the electrodes gradually move apart/closer together all atoms of the central cluster relax in simulated annealing simulations generating a quasi-adiabatic path between the open and the closed conformation.

Not surprisingly the junction rips apart at some finite displacement from the equilibrium, an effect also seen in break-junction experiments. For most junctions this process is accompanied by a surface reorganization on at least one, but often both tips of the electrode(s). When we reverse the process, some junctions snap into the original atomistic conformation (see Fig 3(b)) with subatomic precision(see movies 1-3 in the supplemental material). At the end of the switching simulation we compare the final and the starting geometry. If after the first switching cycle the junction has returned to the same geometry we consider the junction "switchable" and perform further switching cycle simulations to test stability. Otherwise we discard the junction completely and start from scratch.

We have performed 15280 full deposition simulations generating $N_{conf}$=17, 8, 3, 17, 6 junctions with p=1,…, 5 conductance quanta, respectively. Most deposition simulations fail to generate a switchable junction because the "acceptance criterion" for "switchability" was very strict (see methods). We note that the same holds true for most control simulations starting from the "perfect" conformations of Ref [8], indicating that simple rupture of even nearly ideal junctions cannot be the basis of the switching mechanism.

We then compute the zero-bias conductance[22-24] of the entire junction using a material-specific, single particle Hamiltonian and realistic electrode Green's functions[25]. We use a recursive Greens function method[26, 27], which maps the problem of computing the full device Greens function to the calculation of "principal layer" Greens functions, which drastically reduces the computational effort but maintains the accuracy. The electronic structure is described using an extended Hückel model including s-, p- and d-orbitals for each silver atom (7200 orbitals per junction)[28] in the standard minimal basis set of non-orthogonal Slater type orbitals. We take the influence of the leads into account, by assuming a semi-infinite fcc-lattice for the left and the right reservoir. We compute the material-specific surface Green's functions by applying a decimation technique that exploits the translational symmetry of the semi-infinite contacts[29]. We find that the retained junction conformations (typically comprising 500-800 atoms) have a preselected integer multiple (n) of $G_0$ in close agreement with the experiment. The direct comparison of our atomistic, quantum conductance calculations, using the unaltered conformations from the deposition/switching simulations, with the experimental conductance measurements offers a the strong validation of the geometries generated in our deposition protocol. The observed agreement between computed and measured conductance is impressive, because the conductance of metallic wires it is well known to be strongly dependent on the geometry. The extended Hückel was previously shown to give reasonable predictions[18, 30-32] for the conductance of metal nanowires (containing about 800 atoms) where DFT-like methods[11] would be prohibitively costly

We have repeated this process up to 20 times for each junction (Fig 3b) and observe a "training effect", in which the junction geometries become increasingly stable, alternating between two bistable conformations. When recomputing the zero-bias conductance at the end of switching cycle we find the same value as for the original junction (to within ~0.05 $G_0$). Since these observations result from completely unbiased simulations of junction deposition and switching, they explain the observed reversible switching on the basis of the generation of bistable contact geometries during the deposition cycle. If we consider the tip-atoms at each side of the electrodes in the open junction, the equilibrium geometry of both clusters depends on their environment. In the open junction this environment is defined by the remaining electrode atoms on one side, while in the closed junction the tip-cluster of the other electrode is also present. Our simulations demonstrate the existence of two stable geometries for each cluster in both environmental conditions, respectively. Reversible switching over many cycles is thus explained by reversible tip-reorganization under the influence of the gate potential, similar to induced surface reorganization[33-35]. While the overall structure differs between junctions with the same conductance quantum from one realization to the next (see Fig 2(b) for representative examples), the minimal cross-section that determines the conductance is largely conserved (Fig 2(c)). When comparing the opening process and the closing process of the junction we observe asymmetric conductance curves, in agreement with experiment, resulting from several possible low-energy pathways between the open and closed conformations.

These data rationalize the bistable reconfiguration of the electrode tips as the underlying mechanism of the formation of nano-junctions with predefined levels of quantum conductance. These levels are determined by the available bistable junction-conformations, similar to magic numbers for metal clusters[35], that are most likely material-specific. For silver the observed quantum conductance levels appear to coincide with integer multiples of the conductance quantum. When we form a junction by halting the deposition process at a non-integer multiple of $G_0$ (both experimentally and in simulation), subsequent switching cycles either converge to an integer conductance at a nearby level, or destroy the junction. By snapping into 'magic' bistable conformations, junctions are mechanically and thermally stable at room temperature for long sequences of switching cycles. This process is assisted by the electrochemical environment, but not intrinsically electrochemical: the reproducible switching of large junctions by coordinated dissolution/regrowth of the junction is very unlikely.

The electrochemical fabrication approach[5, 7, 9] permits parallel device deposition between multi-electrode arrays structured by lithography. We have varied the deposition protocol to subsequently deposit two atomic transistors on one and the same substrate chip. In Fig. 4 we demonstrate that these transistors can be operated independently and in parallel in a common electrolyte. Each of the transistors is controlled by its own individually addressable switching potential. This operation of two transistors independently on the same chip constitutes the simplest form of an integrated circuit operating on the atomic scale.

Such devices may be manufactured using conventional, abundant, inexpensive and non-toxic materials and possess extremely nonlinear current-voltage characteristics, desirable in many applications. Their electrode arrays can be deposited with lithography, making devices compatible with existing electronics. Because the switching process is achieved with very small gate potential (mV), the power consumption of such devices may be orders of magnitude lower than that of conventional semiconductor-based electronics. Integrated circuits based on this novel principle of operation represent a completely new class of quantum electronic devices, also opening intriguing technological perspectives.

**Acknowledgements**

This work was supported by the Deutsche Forschungsgemeinschaft within the Center for Functional Nanostructures (CFN), Projects C5.1/B2.3 and by grant WE 1863/15-1; we acknowledge the use of the computational facilities at the Computational Science Center at KIST, Seoul.

**Figure Captions**

FIG.1. Switching current by electrochemical, gate-controlled atomic movement. (a) Schematic of the experimental setup: A silver point contact is deposited electrochemically in a narrow gap between two gold electrodes on a glass substrate. Repeated computer-controlled electrochemical cycling permits fabrication of bistable atomic-scale quantum conductance switches. (b) Experimental realization of switching current reproducibly with a single silver atom point contact between an on-state at 1 $G_0$ (1 $G_0$= 2$e^2$/h) and a non-conducting off-state. The source-drain conductance ($G_{SD}$) of the atomic switch (lower diagram) is directly controlled by the gate potential ($U_G$) (upper diagram). (c) Demonstration of quantum conductance switching between a non-conducting "off-state" and a preselected quantized "on-state" at 4 $G_0$. A conductance level can be kept stable, if the $U_G$ is kept at a "hold" level (see arrows).

FIG.2. Relation between the structures of atomic point contacts and their conductance. (a) Quantum conductance switching between a non-conducting "off-state" and a preselected quantized "on-state" at 1 $G_0$, 2 $G_0$, 3 $G_0$, 4 $G_0$, 5 $G_0$, respectively (note individual time axis). (b) Representative conformations of simulated junctions, computed zero-bias conduction and number of junctions with the specified conductance. (c) Representative minimal cross-sections for each conductance level. The minimal cross-sections are characteristic for each group of the switch conformers and determine their quantized conductance.

FIG.3. Simulation of atomic point contact growth and switching process. (a) Snapshots of the deposition simulation. Upper row: The growth process starts with two disconnected Ag (111) layers and stops, when a non-overlapping pathway with a predefined number of silver atoms connects the electrodes. Lower row: Simulation of the switching process reveals bi-stable tip-reconstruction process as the mechanism underlying the reproducible switching of the conductance. During the simulation we kept the grey marked silver atoms at their positions at the end of the deposition and permitted the central cluster to evolve (blue and red atoms) under the influence of the electrochemical pressure. The central silver atoms (red) define the minimal cross section (see Fig 2(c)). These atoms return with sub-Angstrom precision to their original positions at the end of the switching cycle. (b) Difference in the computed conductance between subsequent "on" conformations as a function of the switching cycle for selected junctions of 1 $G_0$, 2 $G_0$, 3 $G_0$, 4 $G_0$, 5 $G_0$, respectively. Junctions switch reproducibly for over 20 cycles between increasingly stable on- and off-conformations (training effect). (c) Variation of the computed conductance of a 5 $G_0$ - switch during one "open-close" process. In agreement with the experimental observations, we find asymmetric plateaus in the conductance curve, if the switch is opened or closed. This can be traced back to the existence of several low energy path ways connecting the open and closed state.

FIG. 4. Parallel and *independent* operation of two atomic transistors grown on one and the same substrate chip in a common electrolyte.

**Supplementary Online Material**

**Methods**

Experimental: The potentials of the working electrodes with respect to the reference and counter electrodes are set by a computer-controlled bipotentiostat. A silver wire (0.25 mm diameter, 99.9985% purity) is used for the counter electrode and the quasi-reference electrode (gate electrode). The electrolyte solution consists of 1mM $AgNO_3$ + 0.1 M $HNO_3$ in bi-distilled water.

Deposition Simulation: The start conformation consists of two parallel Ag(111) layers 38.8 Å apart, with a Ag-Ag distance of 2.88 Å. A Metropolis Monte Carlo method is used to optimize the geometry under the influence of the electrostatic field, calculated by solvinf the Poisson equation for the source charges of the present electrode conformation and a Gupta potential. Each deposition simulation comprises 10000 steps of a random spatial displacement drawn from a uniform distribution of at most 0.1 Å in each direction.

Switching Simulation: During the opening/closing process the silver clusters are displaced in steps of 0.15 Å. For each cluster displacement we perform 10000 simulated annealing steps, where the temperature is reduced from 300 K to 3 K. The initial switching cycle is considered successful, if all atoms return to their original positions to within 0.28 Å.

**Movies of the operating principle:**

1. Switching process of a single point contact based on a *bistable position of a single silver atom*

2. first part: deposition of a silver nano-junction containing five silver atoms in the minimal cross-section
   second part: switching process of this conformation based on a *bistable tip reconstruction*

Figure 1

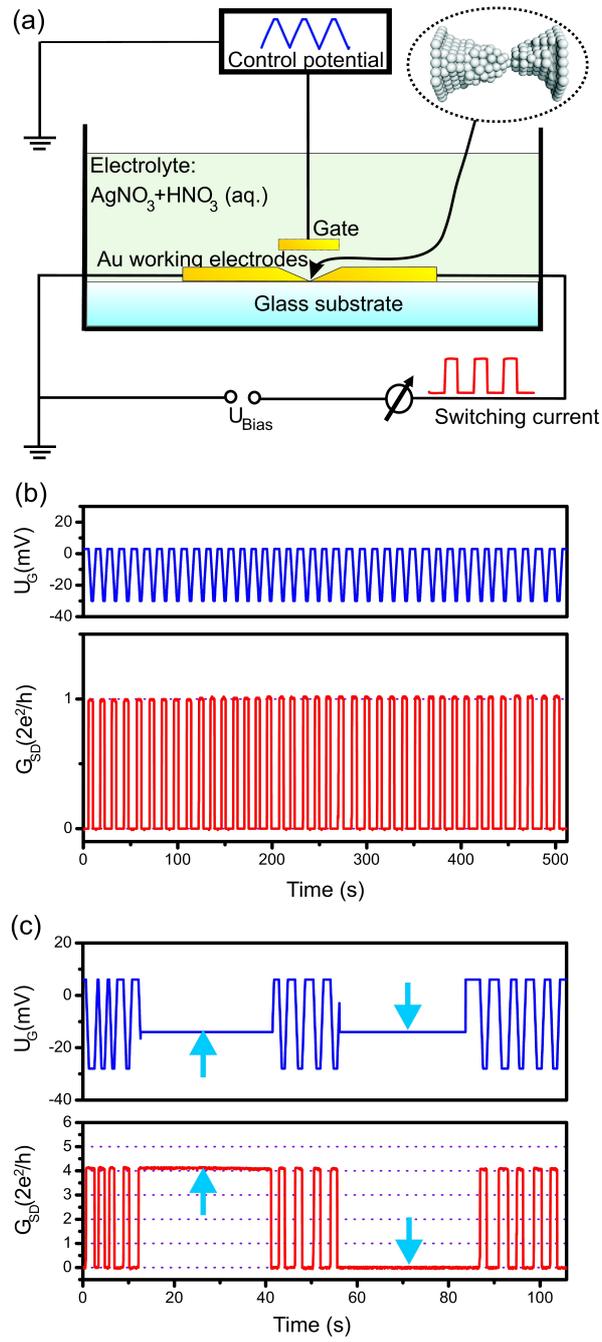

Figure 2

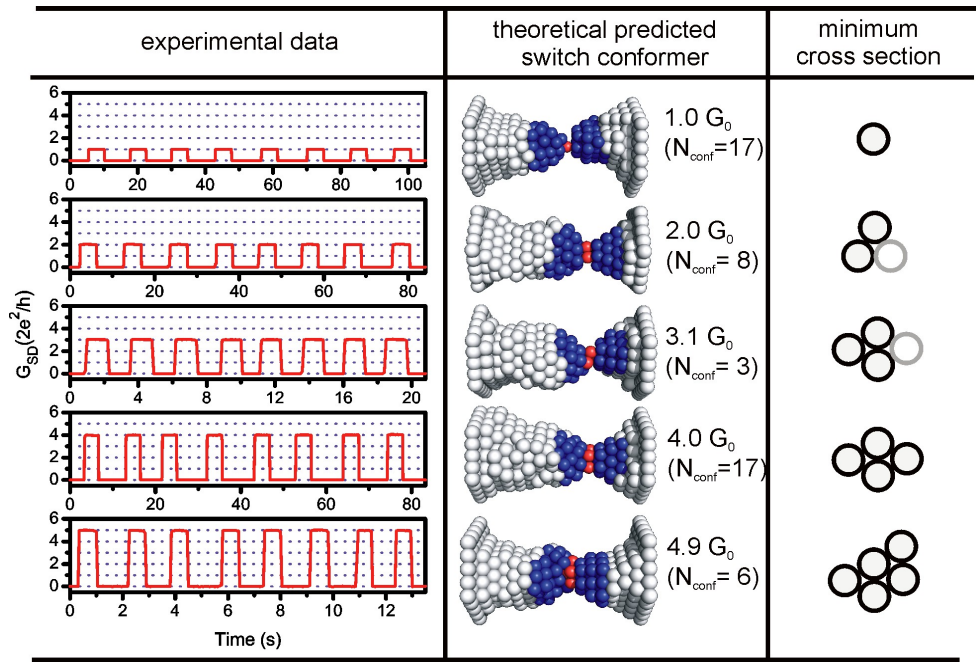

| experimental data | theoretical predicted switch conformer | minimum cross section |
|---|---|---|

Figure 3

a)

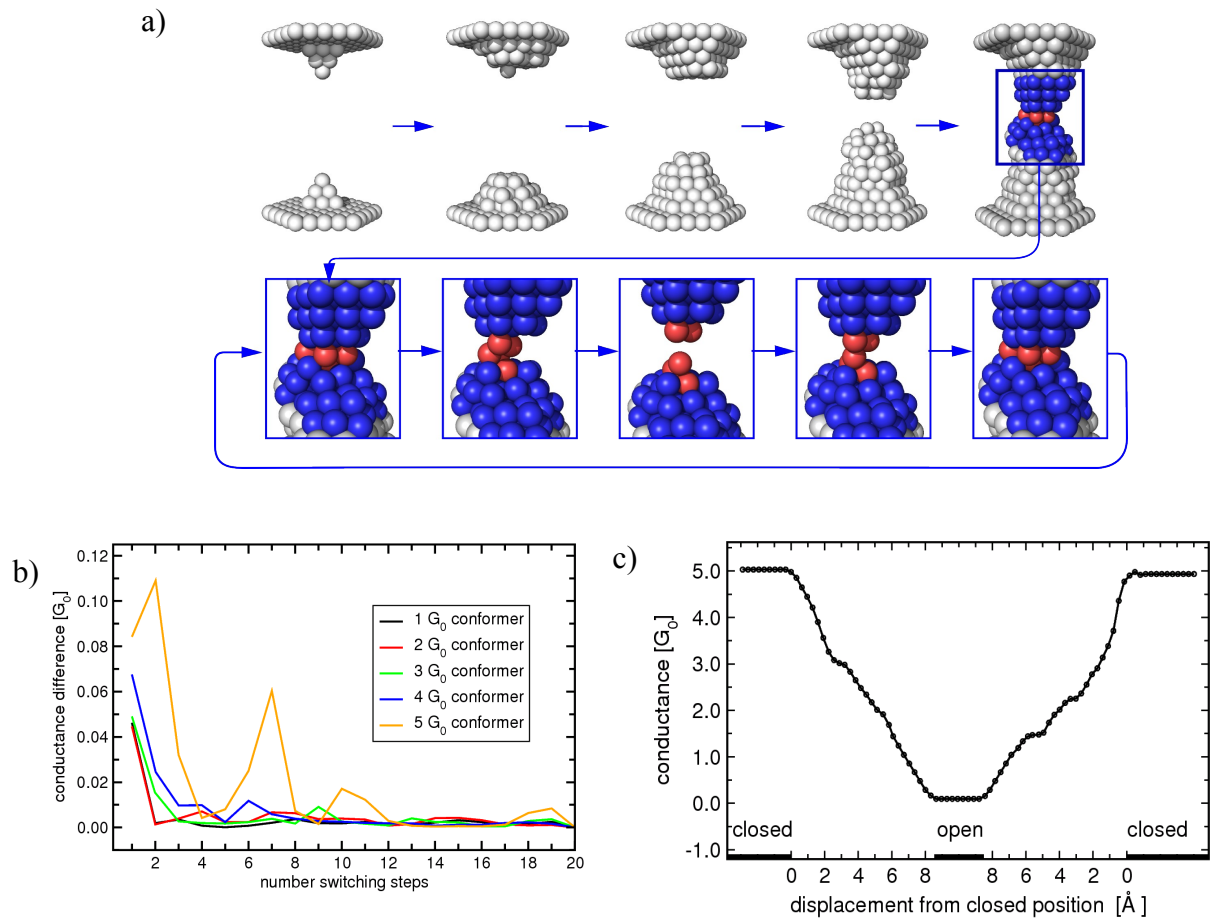

b) [plot: conductance difference [$G_0$] vs number switching steps, with legend: 1 $G_0$ conformer, 2 $G_0$ conformer, 3 $G_0$ conformer, 4 $G_0$ conformer, 5 $G_0$ conformer]

c) [plot: conductance [$G_0$] vs displacement from closed position [Å], labeled closed, open, closed]

Figure 4

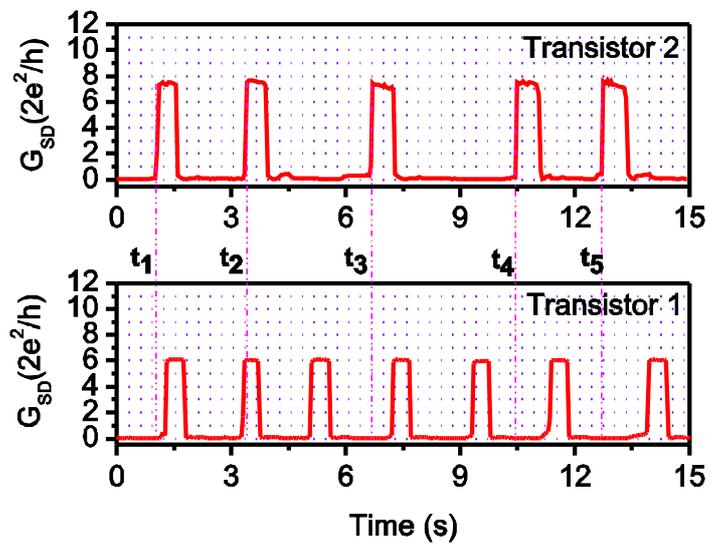

- 17 -